# Progress in the Development of a S-RETGEM-Based Detector for an Early Forest Fire Warning System

**G. Charpak**[a], **P. Benaben**[a], **P. Breuil**[a], **P. Martinengo**[b], **E. Nappi**[c], **V. Peskov**[b,d]

[a] *Ecole Superior des Mines, St.-Etienne, France,*

[b] *CERN, Geneva, CH-1211 Switzerland,*

[c] *INFN Bari, Bari, Italy*

*E-mail*: vladimir.peskov@cern.ch

**Abstract:** In this paper we present a prototype of a Strip Resistive Thick GEM (S-RETGEM) photosensitive gaseous detector filled with Ne and ethylferrocene (EF) vapours at a total pressure of 1 atm for an early forest fire detection system. Tests show that it is one hundred times more sensitive than the best commercial ultraviolet (UV) flame detectors; and therefore, it is able to reliably detect a flame of ~1.5x1.5x1.5 m$^3$ at a distance of about ~1km. An additional and unique feature of this detector is its imaging capability, which in combination with other techniques, may significantly reduce false fire alarms when operating in an automatic mode.

Preliminary results conducted with air-filled photosensitive gaseous detectors are also presented. The approach's main advantages include both the simplicity of manufacturing and affordability of construction materials such as plastics and glues specifically reducing detector production cost. The sensitivity of these air-filled detectors at certain conditions may be as high as those filled with Ne and EF. Long-term test results of such sealed detectors indicate a significant progress in this direction.

We believe that our detectors utilized in addition to other flame and smoke sensors will exceptionally increase the sensitivity of forest fire detection systems. Our future efforts will be focused on attempts to commercialize such detectors utilizing our aforementioned findings.

KEYWORDS: Flame detection; Gaseous detectors; Resistive GEM.

**Contents :**

I.     Introduction

Annually violent forest and bush fires in southern Europe, Australia, the western United States of America and other locations worldwide threaten human lives, decimate industrial and agricultural infrastructures and damage the environment. For example, Table 1 presents worldwide burnt-area data occurring between 1979 and 2000 [1]. Unfortunately, systematic observations indicate the probability of forest and field fires is steadily increasing owing to climate changes and human activities. Considering the worldwide enormity of forest fire destruction, this topic must be acknowledged as a planetary disaster. The most efficient and effective way to fight forest fires is its early identification. Techniques are currently implemented for early fire detection; these can be roughly divided into two main categories: visual smoke and flame detection, and flame infrared radiation (IR) detection. Visual smoke detection operates from watchtowers, small pilotless airplanes ("drones") and even satellites [2]. IR detectors are usually installed in some small patrol airplanes [3]. Other methods are also in development including laser light scattering on smoke and local temperature monitoring with heat detectors [4]. These methods are expensive and offer limited capability. Moreover, since most of the present systems operate in automatic monitoring mode a substantial number of false alarms occur. A combination of several detectors are frequently used in industrial applications such as fire monitoring in small areas to reduce the probability of false alarms, triggered by welding lightening and cetera. For example, IR and UV detectors integrated as one detection system work simultaneously so that a warning alarm is issued only if the signals from both detectors are activated [5].

| Country | Time period | Average number of fires | Total area burned, ha |
|---|---|---|---|
| Albania | 1981-2000 | 667 | 21456 |
| Algeria | 1979-1997 | 812 | 37037 |
| Bulgaria | 1978-1990<br>1991-2000 | 95<br>318 | 572<br>11242 |
| Cyprus | 1991-1999 | 20 | 777 |
| Croatia | 1990-1997 | 259 | 10000 |
| France | 1991-2000 | 5589 | 17832 |
| **Greece** | **1990-2000** | **4502** | **55988** |
| Israel | 1990-1997 | 959 | 5984 |
| Italy | 1990-1999 | 111163 | 118576 |
| Lebanon | 1996-1999 | 147 | 2129 |
| Morocco | 1960-1999 | n.a. | 2856 |
| Portugal | 1990-1997 | 20019 | 97175 |
| Romania | 1990-1997 | 102 | 355 |
| Slovenia | 1991-1996 | 89 | 643 |
| Spain | 1990-1999 | 18105 | 159935 |
| Turkey | 1990-1997 | 1973 | 11696 |

**Table1.** Fire statistical data of the Mediterranean countries and its neighbors in South Europe [1].

In this paper, based on our early developments [6], we advocate applying the same approach to forest fire detection and enforcing existing monitoring systems with the UV flame detection. This will prove beneficial for automatic systems, which are exclusively applicable in large area forest surveys, in combination with other detectors as well as for monitoring the areas where fire has the highest probability to start: glades; bushes; and dry areas with small trees.

Note that there are commercial UV flame detectors, and the most sensitive among them is the so–called **"EN 54-10 class- 1"** [*] which can detect ~30x30x30cm3 flame from a distance of ~20m in 20sec. Its sensitivity, however, is insufficient for forest fire detection from a distance of more than 200-300m (see Paragraph III. 2).

In contrast to existing commercial detectors, our UV flame detector operates on a principle of photoionization of vapors of low ionization potential (such as EF [8] or TMAE [8]). This approach has several advantages and, among them, the most important is an exceptionally high sensitivity. Depending on their design, the proposed flame detectors are one hundred to a thousand times more sensitive to the UV radiation produced by a flame than first class commercial detectors. This paper describes the latest progress in the development of gaseous detectors for forest fire detection. Its primary focus is the operation of the S-RETGEM, which allows for constructing a simple, rather inexpensive and robust detector featuring imaging capability, which, used in conjunction with other methods, will considerably increase forest fire detection efficiency and reduces false alarm occurrences.

---

[*] The sensitivity of the class -1 UV detectors is comparable to the sensitivity of the best commercial IR detectors [7].

## II. Detectors Design

Three gas chambers identical in designs, but made of different materials: stainless steal (SS), Al and Plexiglas-were used in these studies (see Figure 1).

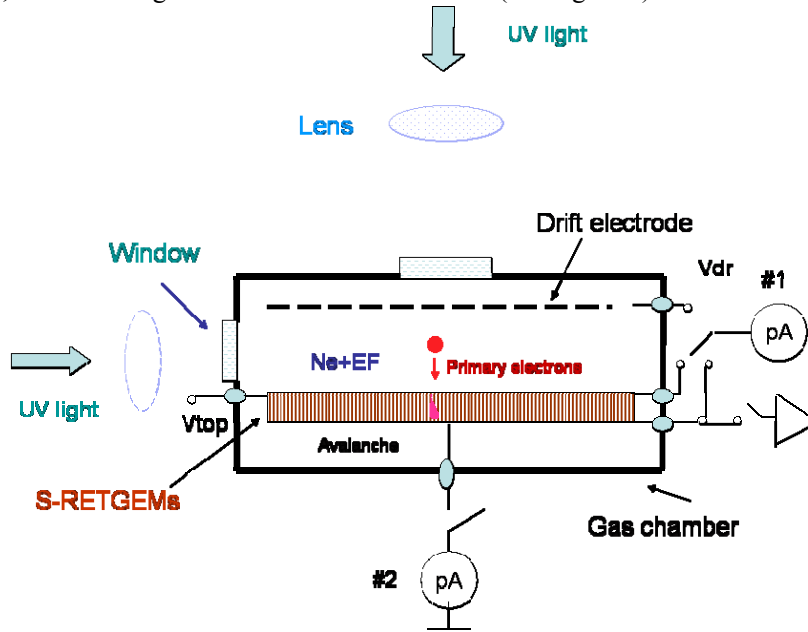

**Figure 1**. A schematic drawing of the experimental setup

Comparative measurements taken with gas chambers allowed us to make the best choice and recommendation for the final commercial prototype. These gas chambers have a cylindrical shape with dimensions of 17 cm diameter and 10 cm height. Each of these has two quartz windows: one on the top, 40 mm in diameter, and one on the side of the cylinder, 20-30 mm in diameter. Inside each gas chamber, an S-RETGEM was installed with a drift mesh typically placed 2 cm above it. The concept and the design of the S-RETGEM are described in several recent papers [9, 10]. Recalling the main features, it is a Thick GEM [11-13] with double-layered micro pattern electrodes with an inner layer of thin Cu strips and an outer layer of a resistive grid, manufactured by screen printing technology, on top of Cu strips. A magnified photo of this structure is shown in **Figure 2.**

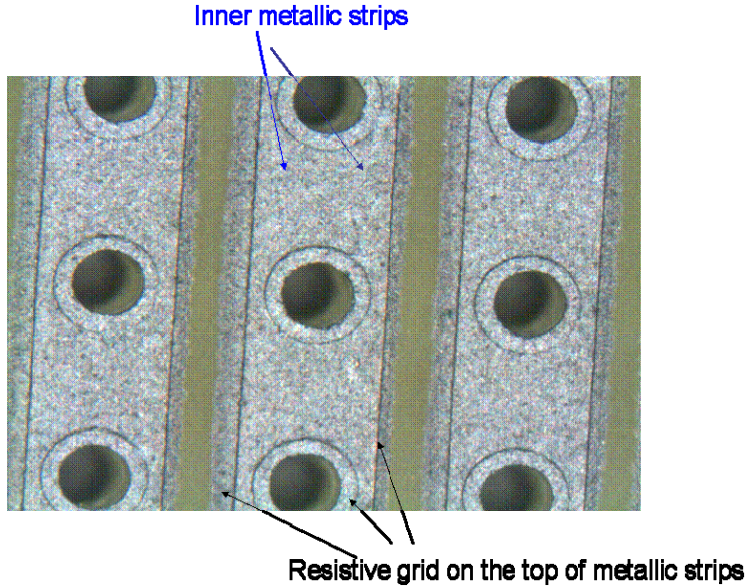

**Figure 2.** A magnified photo of the S-RETGEM. One can see the resistive grid and the inner metallic strips

Two slightly different designs of S- RETGEM were used in this work: one has a thickness of t=0.4 mm, a holes diameter of d=0.3mm, pitch p=0.8mm; and a strip width of w=0.4mm, the other one has t=0.8 mm, d=0.4mm, p=1mm, w=0.5mm. Both designs have an active area A of $10 \times 10 cm^2$ and strips on one side orient perpendicular to the strip on the other side which offers a possibility of a two dimensional (2-D signal readout [14]) Note that S-RETGEM has several important advantages over the earlier version with un-segmented resistive electrodes named RETGEM (also used for the UV imaging purposes [6]). First, it is more robust due to the low capacity of strips; the energy released in occasional sparks is at least five times less than in RETGEM. Regarding the problem of persistent sparking at gas gain below the nominal, one can electronically disconnect the strip to stop the sparking or locally lower the voltage in this detector region so that the rest of the detector remains in operational condition. Second, given its low strip capacity, it provides a higher signal to noise ratio than RETGEM. Third, by measuring signals from each strip, one can determine a 2-D position of the created photoelectrons with an accuracy of a few mm, eliminating the necessity of an additional readout plate usually used for the purpose [6]. Fourth, S-RETGEM has a high ratio of active area A to the total area S: A/S~90% whereas in the case of the RETGEM it was only 60%. All this allows utilization of just a single S-RETGEM (instead of a triple element detector previously employed [6]: a double RETGEM and a signal readout plate).

The gases used in this work were Ne+EF, Ne+10%$CH_4$+EF, Ne+EF +air (various percentages) and pure ambient air saturated with EF vapors. Differing from the previous work [6] the EF vapors were introduced into the gas chamber by flushing carrier gas for about 30 sec through the bubbler filled with liquid EF and heated to 90°C. As will be subsequently revealed (see paragraph III.1d) this offers higher sensitivity at elevated ambient temperatures. Before filling with the gas mixture, the SS and the Al chamber were pumped for several days at vacuum of $10^{-6}$ Torr and heated to a temperature of 80°C. Next, the chamber was cooled to room temperature; the gas mixture was introduced; and the chamber was sealed. The Plexiglas chamber, given its small wall thickness, cannot be pumped and was simply flushed for a few days with the selected gas mixture (typically Ne +air +EF). Lastly, it was sealed. The tests were performed with various flames consisting of differing sizes and functional objects (candles, cigarette lighters, gasoline flames, and wooden flames). Additionally, in some measurements a Hg UV lamp was used (often combined with a narrow-band filter having a peak of transmission at 185 nm), as well as various radioactive sources: $^{55}$Fe, $^{90}$Sr, and $^{241}$Am. Frequently, all S-RETGEM strips on each detector side were connected together and one side was biased by high voltage (HV) whereas the other side was connected to a charge sensitive amplifier Ortec 142PC. For imaging purpose 10 or 15 selected strips (on which the light was projected by a focusing lens -see Figure 1 and [6]) were each was connected to custom-made charge-sensitive preamplifiers; further, remaining strips were grounded via 10MΩ resistor. Most of the measurements were conducted in counting mode, and the avalanche-induced pulses from the S-RETGEM were electronically recorded and counted. However, some measurements (especially for the gas gain calibration) were performed in current mode in which photocurrents produced by an Hg lamp were measured with picoammeters #1 and 2 (see Figure 1)

In the comparative studies we used both class-1 Hamamatsu R2868 flame detector and a custom made single- wire TMAE filled detector [15].

### III. Results
#### III.1 Results of Laboratory Tests
III.1a) Gain Measurements with a Hg Lamp and Radioactive Sources

As previously noted, the gas gain measurements were performed in current mode. For example, Figure 3 illustrates the current (curve #1) measured from the picoammeter #1 as a function of the negative voltage applied to the drift mesh $V_{dr}$ where a filtered UV radiation (185nm) from the Hg lamp entered the side window of the gas chamber. To ensure complete photoelectron collection in these measurements the top and bottom electrodes of S-RETGEM were connected together to the picoammeter. This photocurrent current initially increases sharply with $V_{dr}$ and then at voltages $V_{dr}>V_{sat}$ reaching a saturated value $I_{sat}$.

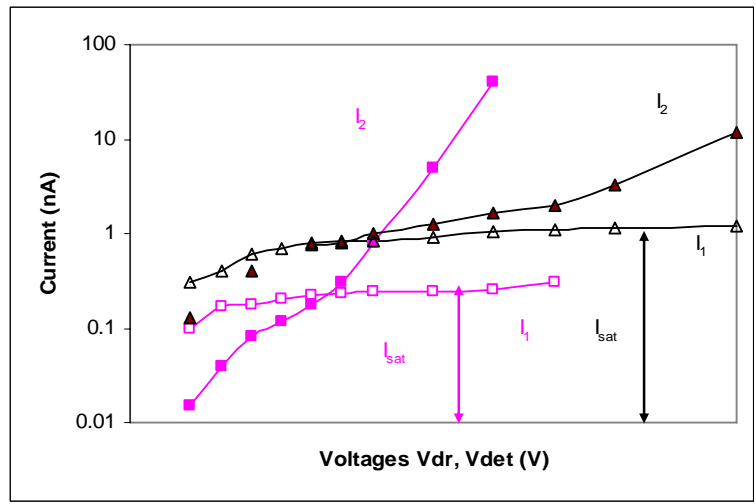

**Figure 3.** Currents measured from the top S-RETGEM electrode (open symbols) and from the bottom electrode (filled symbols) as a function of $V_{dr}$ and $V_{det}$ respectively. In rose color are marked the results obtained in Ne+EF (S-RETGEM 0.8mm thick), whereas in black color-the results obtained in Ne+10CH$_4$+EF (0.4mm thick). The saturated values of the photocurrent $I_{sat}$ are also marked in the figure.

Usually at the plateau region of the current almost all electrons and ions created by the ionizing radiation escaping mutual recombination and collecting on electrodes so that for the monochromatic UV radiation with the wavelength $\lambda$:

$$I_{sat}=b\,L(\lambda)Q(\lambda)[\,1-\exp\{-k(\lambda)\sqrt{A}\}]\quad (1),$$

where b is a coefficient, $L(\lambda)$ is the UV lamp radiation intensity at the wavelength $\lambda$, $Q(\lambda)$ is EF vapors quantum efficiency and $k(\lambda,T)$ is the EF absorption coefficient at given temperature T.

In the next set of measurements the current was recorded from the bottom S-RETGEM electrode-assisted by picoammeter #2-as a function of the negative voltage applied to the top electrode $V_{det}$ ( or in other words as a function of the voltage applied across the S-RETGEM). In this case the voltage drop between the drift mesh and the top electrode was kept constant:

$$V_{dr}-V_{det}=const=V_{sat}\quad (2).$$

Discernible from Figure 3, this current $I_2$ (curve #2) significantly increases as a function of the voltage $V_{det}$ and reaches very high values so in measurements at high $V_{det}$ UV attenuators were employed to gradually reduce the intensity of the radiation from the Hg lamp on orders of magnitude. The efficient gas gain of the S-RETGEM was defined as:

$$G=I_2/I_{sat}\quad (3).$$

Next, **Figure 4** exhibits gain curves $G=G(V_{det})$ measured in Ne+EF for 0.4 and 0.8 mm thick for S-RETGEMs. Values of the gas gains achieved were remarkably high, $G>10^5$, without any sparking. One unique feature of Ne-based mixtures is that they produce approximately one hundred times higher gas gains as compared to other mixtures in particular Ar+EF [9, 16]. In the presence of X-rays from $^{55}$Fe or electrons from $^{90}$Sr (modeling cosmic muons) the maximum achievable gain was around $10^5$ sufficient for single electron detection where the measurements were performed in counting mode. This is an additional, singular feature of Ne-based mixtures whereas in other gases such as Ar -based mixtures the maximum gains achieved with radioactive sources are usually one hundred to one thousand times less than with the UV light [ 9,16]. As a result in Ar-based mixtures one cannot use a single-step detector for recording UV light in pulse counting mode in the presence of the radioactive background [6].

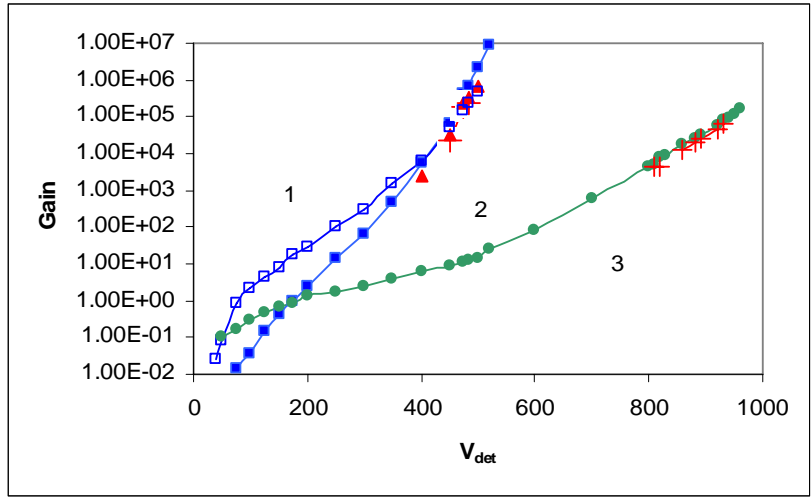

**Figure 4.** Gain cuves mesured with SRETGEM in current mode (UV) and pulse mode (Fe and Ru):
1)Ne+EF, S-RETGEM thickness t=0.4 mm, 2) Ne+EF, t=0.8, 3) Ne+10%CH$_4$, t=0.4 mm.
Red symbols represent the results obtained in counting mode with $^{55}$Fe (triangles) and $^{90}$Sr (crosses).

III.1.b) Small Flames Tests

As was mentioned above, as main advantage, we could easily detect UV flame emission with single S-RETGEM in photon counting mode as the gas gain Ne+EF even in the presence such radioactive sources as $^{55}$Fe and $^{90}$Sr. At $G\sim10^5$ only alpha particles from $^{241}$Am, which model very rare events caused by neutrons, triggered sparks, yet the energy remained negligibly small owing to resistive protective layers, low strip capacity and low operational voltages typical for Ne -based mixtures. These sparks never damaged either the S-RETGEM or the front-end electronics.

In Figure 5 depicted results of photon counting rate measurements involving a candle positioned at distances of three, five, ten and thirty meters from our flame detector. The figure also displays a dash line representing the calculated counting rate given the assumption that the light intensity from the candle obeys $1/R^2$ dependence where R is a distance between the flame and the detector. The measured and calculated values well coincide indicating unnoticeable absorption of the flame UV light at these distances; otherwise; the measured counting rate should be below the calculated one. A comparison with similar measurements performed with the Hamamatsu R2868 detector shows that it is one hundred times less sensitive than our S-RETGEM. Figure 6 gives counting rates from the S-RETGEM detector obtained under varying circumstances: in a room with an open window on a sunny day; outside the room exposed to direct sun irradiation; and outside the room exposed to indirect sun irradiation. Clearly when the S-RETGEM was directly irradiated by the Sun, the background counting rate only slightly increased (about a factor of two) above the natural cosmic background. The detector easily detected the candle in the same solid angle as the direct sunlight given this low sun-induced setting.

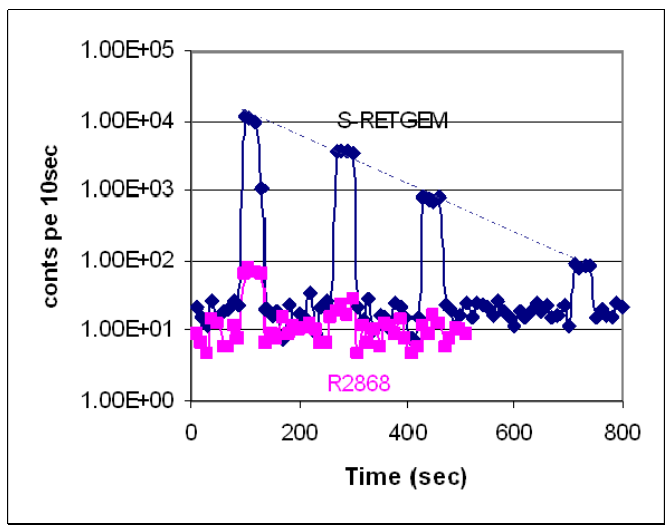

**Figure.5.** Counting rate measured from S-RETGEM and Hamamatsu R2868 vs. time at various conditions. At time ~100 sec, ~300sec, ~450sec and ~700sec, both detectors were exposed to the candle light which was places at different distances (indicated by numbers in the figure) from the detectors. In between the measurements both detectors recorded a background counting rate. A dash line indicates calculated counting rate assuming that it obeys $1/R^2$ dependence.

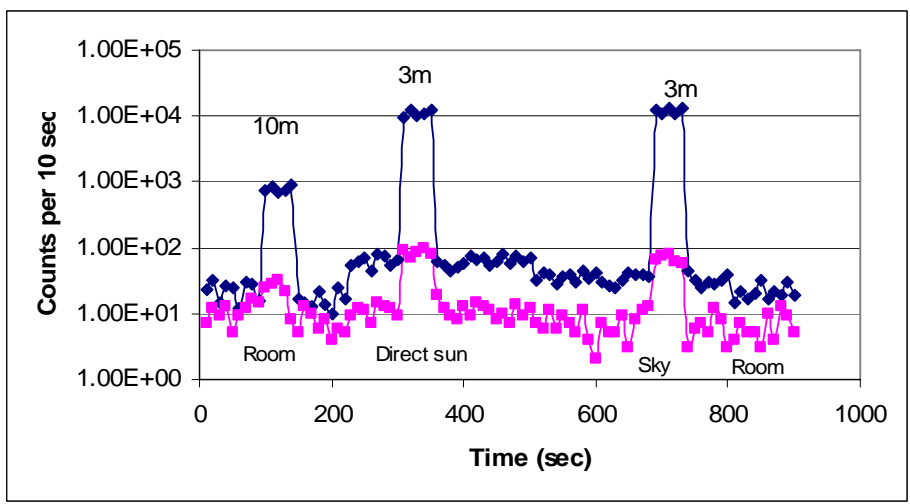

**Figure 6.** Counting rate measured from S-RETGEM and Hamamatsu R2868 vs. time at a sunny day at the following conditions: inside the room with the window open (indicated "room"), outside the room exposed to the direct sun irradiation ("direct sun"); and outside the room exposed to indirect sun irradiation ("sky")

III.1.c) Long-Term Stability Tests

We conducted the crucial task of evaluating the long-term stability of S-RETGEMS placed in sealed gas chambers consisting of varying materials, specifically SS, AL and Plexiglas. Figure 7 shows peculiar findings obtained with Ne+EF and Ne+10%$CH_4$+EF as well as results derived from Ne+EF + air. It must be underlined that gain degradation remained negligible with the SS chamber. Within the Al chamber, the gas gain slightly degraded in Ne+EF gas mixture. The stability achieved proved appreciatively better with Ne+10%$CH_4$+EF indicating that in the presence of the quenching gas ($CH_4$) the effect of the chamber outgassing on gas gain was considerably reduced. The worse foreseeable results were obtained with the Plexiglas chamber attributable to the poor quality of this chamber, a small gas leak and the out-gassing nature of Plexiglas material. Improved stability was achieved by adding $CH_4$. Although it poses a potential cost liability, the sealed gas chambers should be made of SS. Part III.3 focuses on Plexiglas chamber demonstrations indicating remarkable stability gains when varying air percentages are added to the Ne+EF mixture. This information offers ideas concerning the viability of manufacturing affordable plastic gas chambers housing S-RETGEMS.

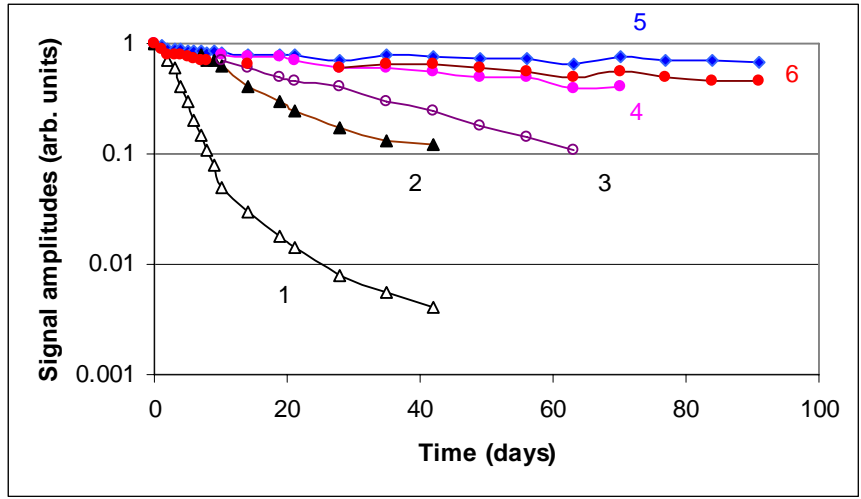

**Figure 7.** Gas gain variation (arbitrary units) vs. time measured with three chambers using $^{55}$Fe source: 1) Plexiglas chamber filled with Ne+EF; 2) Plexiglas chamber, Ne+$CH_4$+EF; 3) Al chamber, Ne; 4) Al chamber Ne+$CH_4$+EF; 5) SS, Ne+EF; 6) Plexiglas chamber, Ne +EF+ 25%air.

III.1d) Temperature Stability Tests

Because an excess of the EF vapors was initially introduced in to the gas chambers (the bubbler was kept at 90°C so that at this temperature vapors condensed inside the chamber and formed a thin adsorbed liquid layer on all inner surfaces) the EF vapor pressure increased with ambient temperature. As a results, at the fixed operational voltage ($V_{det}$=const) the pulse amplitude consequently increased with the temperature too as well as a S-RETGEM detection efficiency (note that Ne+EF is a Penning mixture so in some temperature interval the gas gain increases with the EF vapors pressure).

At the fixed flame intensity and the fixed distance between the flame and the detector (R=const) the recorded number of pulses is:

$$N(T) = c \int F(\lambda) Q(\lambda) [1 - \exp\{-k(\lambda,T)\sqrt{A}\}] \quad (4).$$

where $F(\lambda)$ represents the emission spectrum of the flame and c is a coefficient.

Data derived from S-RETGEM efficiency versus the temperature or precisely N(T) are presented in Figure 8. These measurements were conducted under two conditions: first, the UV flame radiation entered the gas chamber through the side window; second, it entered through the top window. The same figure distinguishes dash lines representing calculation results based on formula (4). These calculations and measurements proved tenable except under two conditions: low temperatures such as T<300K; and light introduced through the top window. This discrepancy arises as a consequence of the contribution EF absorbed layer condensed on the top

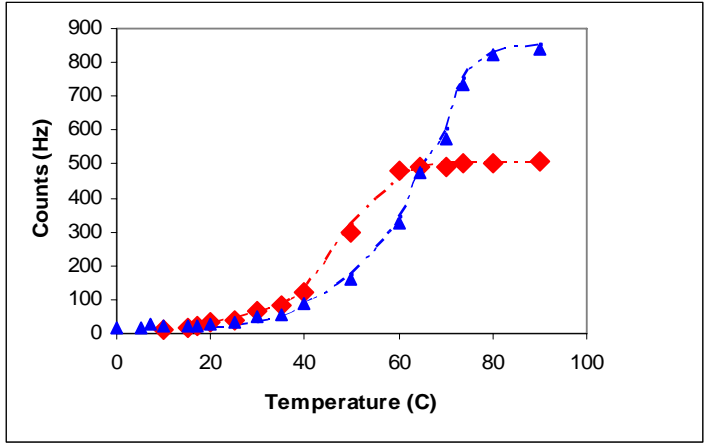

**Figure 8**. Counting rate measured from S-RETGEM irradiated parallel to the S-RETGEM (red symbols) and perpendicular to it (blue symbols). Note that in the later case due to the surface effect (photosensitivity of the EF absorbed layer [15] the sensitivity in the temperature interval 0-30°C drops compared to 30°C only on a factor of 2. At elevated temperatures (the possible situation when the detector is warmed up by the direct sun radiation in a hot summer day) the sensitivity increases 6-10 times

electrode of the S-RETGEM (see [15,17] for details). An examination of the results presented in [6] indicates that adding of the excess of EF to the chamber produces a higher sensitivity compared to the responsiveness of the chamber filled with the saturated pressure of the EF at room temperature. Note that forest fires usually appear at dry and very warm weather, so it is very favorable that the sensitivity of the S-RETGEM at $V_{det}$=const increases with the temperature.

### III.2. Results of Tests Performed at Long Distances

The S-RETGEM measurements outlined in previous paragraphs were performed at distances R≤30m and without discernable UV flame absorption. But quite importantly, S-RETGEM must demonstrate the ability to record flames at greater distances in order to present a realistic application of forest fire detection. Relevant tests were undertaken simultaneously with three detectors: S-RETGEM; TMAE- filled single-wire counter [15]; and Hamamatsu R2868 Uvitron.

Figure 9 illustrates certain measurement results. We experimented with a wooden fire (an approximate size 1,5x1,5 x1,5 m$^3$) on a sunny day and recorded the first set of measurements at R=50 m (a reference point) without an expectation of strong light absorption. We gradually lengthened the distance to 800m (the maximum size of the area available for these measurements). Open symbols in this figure indicate the calculated counting rate vs. distance with the assumption that it obeys 1/R$^2$ low. It is evident that both S-RETGEM and the TMAE – filed counter efficiently detect the flame at a distance up to 800 m; however, the counting rate deviates from the 1/R$^2$ dependence. This allows for a rough estimate of air absorption happening

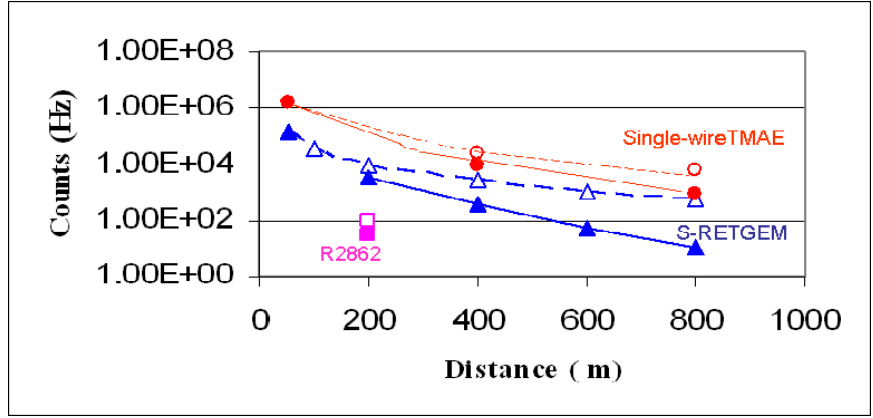

**Figure 9**. Counting rate vs. distance measured with three detectors (SRETGEM, single wire counter filled with TMAE vapors and Hamamatsu R2868) at 27°C from a flame $1.5\times1.5\times1.5\text{m}^3$ in a sunny day

at great distances. Indeed, at this point of light absorption, the counting rate of the detector will be:
$$N(R)=(a/R^2)[\int F(\lambda) QE(\lambda) \exp\{-RC(\lambda)\} )\{1-\exp\{-k(\lambda,T)\sqrt{A})\} \, d\lambda], \quad (5).$$
where $C(\lambda)$ is the light absorption coefficient in air and a is a coefficient depending on the position of the reference point.

Numerical calculations of the convolution
$$\int F(\lambda) QE(\lambda) \exp\{-RC(\lambda)\} )\{1-\exp\{-k(\lambda,T)\sqrt{A})\} \, d\lambda \quad (6)$$
indicate that the detector response function is sharply peaked at $\lambda_1\sim 195$ nm for EF and $\lambda_2\sim 215$ nm for TMAE. In a few words this is because the air sharply cuts the radiation in the interval
$\lambda <185$ nm for short distances and $\lambda <220$ nm at longer distances, whereas the $Q(\lambda)$ is zero at $\lambda_{i1}> 200$ nm for EF and $\lambda_{i1}>220$ nm for TMAE and the flame emission steadily increases with the $\lambda$.

Thus for the sake of simplicity in the first approximation, one can replace the integral (5) with simpler formulas:
$$N_{EF}(R)=(ba/R^2) F(\lambda_{i1}) QE(\lambda_{i1}) \exp\{-RC(\lambda_{i1})\} )[1-\exp\{-k(\lambda_{i1},T)\sqrt{A})\}] \quad (7)$$
$$N_{TMAE}(R)=(ba/R^2) F(\lambda_{i2}) QE(\lambda_{i2}) \exp\{-RC(\lambda_{i2})\} )[1-\exp\{-k(\lambda_{i2},T)\sqrt{A}] \quad (8)$$
where $N_{EF}$ and $N_{TMAE}$ are counting rated measured with detectors filled with EF and for TMAE respectively.

From the analysis of the measured at $N_{EF}(R)$ and $N_{TMAE}(R)$ the following valued of the "characteristic" absorption coefficients were determined: $C(\lambda_{i1})\sim 1/250$m and $C(\lambda_{i2})\}=1/400$m.

Applying these absorption coefficients, one can estimate that a $5\times5\times5$ m$^3$ fire can be detected with EF filled S-RETGEM at a distance of 1,5km if the counts are accumulated during 20 sec. By increasing the recording time (the interval between when the fire erupts and when the alarm sounds) to 100 sec respectively higher sensitivity may be achieved.

Note that the Hamamatsu R2868, given its low efficiency, produced very poor results in detecting of the wooden fire (see Fig. 9).

### III.3. Imaging Tests

Due to the complicity of the aforementioned long-distance measurements, we were not able at the same time perform any imaging of flames with our S-RETGEM. This was done in separate measurements. The detector arrangement under which this image was taken was different from the one described in the earlier work [6]. In the quoted work the light focused by the lenses entered the detector through the horizontal window and the signals were measured form the readout plate having a fun out strip geometry. In the present work we did not use a S-RETGEM with the fun out strips, so, we turned the chamber to a 90° such that its axis remained horizontal whereas the S-RETGEM strips were oriented perpendicular to the horizontal plane. Subsequently, the light focused via the lens (see Figure.1) infiltrated the top window, now becoming perpendicular to the horizontal plane, and creating an image on the top electrode of the S-RETGEM. A read-out of strips displays a one-dimensional image of the flame. Refer to Figure 10 to observe a digital image (or the number of counts per strip) of a gasoline flame ~20x20x20 cm$^3$ detected on a sunny day at a distance R=300 from the S-RETGEM. In effect, its critical imaging capability makes the S-RETGEM, an "intelligent" detector capable of efficiently rejecting various false signals. This is especially true if a network of S-RETGEM works to monitor the forest area or if S-RETGEM operates in conjunction with various other flame detectors.

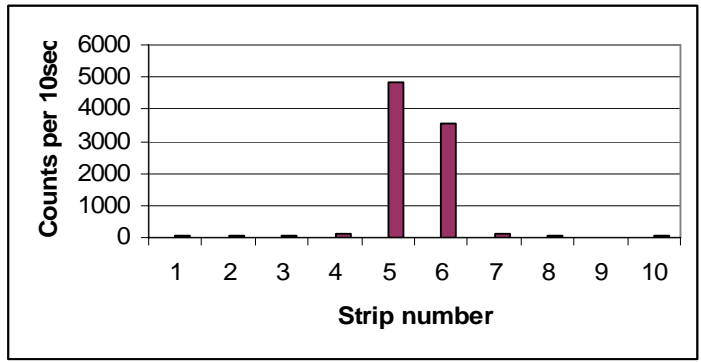

**Figure 10.** A digital image of the gasoline flame ~20x20x20cm$^3$ recorded with a S-RETGEM at a distance of R=300m at temperature of 24°C.

### III. 4. Study of Air- Filled Photosensitivity of S-RETGEMS

In recent papers [18, 19] we outlined how a RETGEM detector stably operates at rather high gas gains in ambient air and detects not only alpha particles, but even soft x-rays. Encouraged by these earlier results, we tested if S-RETGEMS can achieve sufficiently high gas gains in mixtures of Ne with air and EF vapors as well in pure air saturated with EF vapors. Preliminary results of these tests are presented in Figures 11 and 12. In particular Figure 11 cites gas gain curves vs. $V_{det}$. It is discernible that gains ~10$^5$ can be achieved in Ne with low air percentage in the presence of the EF vapors.

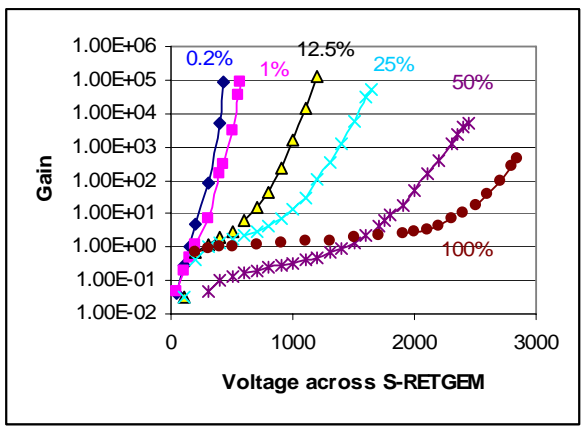

**Figure 11.** Gas gain of S-RETGEM measured in current mode (using UV light from the Hg lamp) in mixtures of Ne with varios air concentration (marked in the figure)

In pure air the "ordinary"[*] S-RETGEM is only capable of reliably operating at gains of below $10^3$.

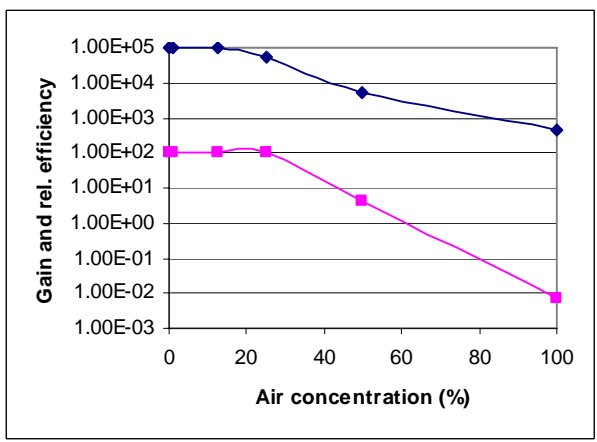

**Figure 12.** The maximum achievable gain (upper curve) and the relative efficiency of S-RETGEM (lower curve) in Ne+ air mixtures vs. air concentration.

Figure 12 plots the maximum achievable gains of "ordinary" (unselected) S-RETGEM versus the air concentration as well as the detectors' efficiency measured with respect to Hamamatsu R2868. As is evident, the air concentration sustained below 50%, the S-RETGEM proved five to hundred times more sensitive than the Hamamatsu R2868. Preliminary long-term stability test results of the S-RETGEM operating in mixtures with air are presented in

---

[*] In 100 % air saturated with the EF vapors the gas gain may reach as high as $10^4$; however, the operational voltage $V_{det}$ becomes rather high (~3kV) thus, only exceptional S-RETGEM withstands this voltage.

Figure 7 and underscores the remarkable stability of the Plexiglas chamber filled with Ne+25% air +EF. More tests need to fully certify this effect. The successful operation of S-RETGEM in air +EF make it possible to dramatically simplify the gas chamber design and thereby reduce its cost by allowing it be manufactured from inexpensive construction materials such as plastics and glues. Such detectors may operate stably notwithstanding minimal out-gassing even in the presence of a minor gas leak. We are enthusiastic about our continued tests on the S-RETGEM given these very promising preliminary findings.

**IV. Conclusions**

Based on our findings, we advertise the new S-RETGEM-based position-sensitive UV detector of flames and sparks to facilitate early forest fire detection. Preliminary investigations confirmed that the sensitivity of this detector is at least one hundred times higher than that of any other commercially available UV flame detectors. The new detector's high sensitivity, sound timing properties and imaging capability qualify it to survey extensive land areas thus making it a cost efficient alternative to the task of several conventional detectors performing the same task.

S-RETGEMs –based detectors can be installed on towers in troubleshooting areas including bushes and open areas and optimally implemented in combination with other types of flame detectors including CCD, IR and more. The most advantageous application involves joint operation with IR detectors in small patrol airplanes since the transmission of the atmosphere rapidly increases with the altitude [20] enabling S-RETGEM detectors to identify small fires at a distance of a few kilometers. S-RETGEM-based flame detectors are additionally suitable for industrial applications widely ranging from factories to aircraft hangars. Finally, Table 1 presents a summarization of the main experimental work results including rough cost estimations pertaining to various detectors. S-RETGEM featuring acute sensitivity can be manufactured at comparable commercial detector prices.

| Detector | Price (Euro) | Sensitivity (Ham. units) |
|---|---|---|
| **Hamamatsul** | **80** [21] | **1** |
| Sealed single wire (TMAE vapors) | 120-150 | 1000 |
| Seaed S-RETGEM (Ne+EF) | 100 [22] | 100 |
| Sealed S-RETGEM, pos-sens, Ne+EF | 150 | 100 |
| S- RETGEM, pos. Sens Ne+ air | 80 [22] | 25 |
| S-W filled with air (see appendix) | 40-70 [21] | 10 |

**Table 2.** Summary of the sensitivities achieved with various detectors tested in this work and preliminary estimation of their production cost

In conclusion, we believe that S-RETGEM based detectors operating in combination with other flame and smoke sensors will powerfully increase the sensitivity of forest fire detection systems. With this in mind, our future efforts will be focused on attempts to promote such detectors for commercial use.

## V. Acknowledgment

Authors would like to thank Ms. M. Louise Nudo for her contribution at the final stage of this work

### VI. Appendix

This paper explores a promising direction in the development of air-filled imaging detectors or S-RETGEMS. Obviously S-RETGEMS filled with 100% ambient air and EF vapors pose effective and promising applications. However, not all S-RETGEMS can withstand the necessary high voltage applied across the detector, and this presents us with a challenge. We are investigating whether higher gains are achievable with a single-wire counter and experimenting with a design of the dielectric supporting structure situated between the cathode cylinder and the anode wire to ensure functioning at very high voltages. Figure 13 presents our design of single-wire counter filed with air and EF vapors. The large size and high quality of the dielectric interface between the anode wire and the cathode cylinder enabled us to apply very high voltages to this detector producing remarkably elevated gas gains. Refer to Figure 14 for examples of signal amplitudes versus the applied voltage measured from this single- wire counter when it detected 6 keV x-rays from the $^{55}$Fe and single photoelectrons produced by the UV light via the photoionization of the EF vapors. Gas gains up to $10^7$ are clearly achieved in this case. The comparative measurements reveal that the sensitivity of this single-wire counter to flames was ~10 times higher than Hamamatsu R2868 capability. Thus, such a detector serves as a practical, alternative option for certain applications.

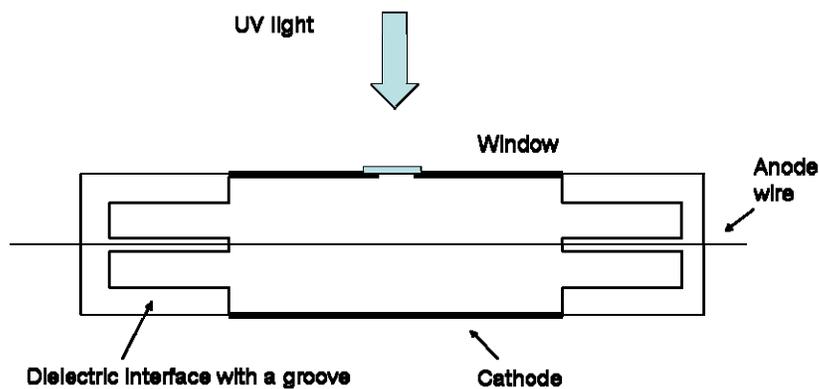

**Figure 13**. A schematic drawing of the single–wire counter capable to operate at high gas gains in mixture of air with saturated EF vapors.

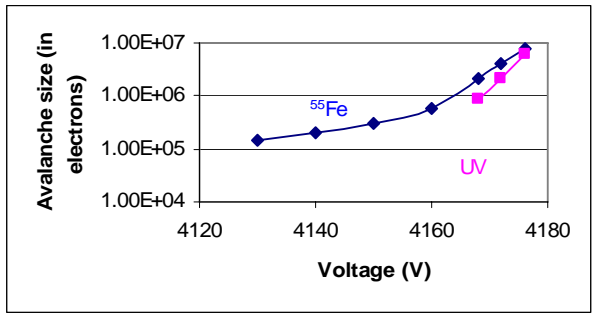

**Figure 14.** Avalanche size in electrons due to multiplication in air saturated with EF vapors at room temperature (22°C). Upper curve is measured in pulse mode with [55]FE source, the lower one measures in pulse mode with the UV light (single photoelectron detection).